\documentclass[12pt]{iopart}

\usepackage{amssymb}
\newcommand{\interior}{\, \lrcorner \,}

\usepackage[applemac]{inputenc}

\begin{document}

\title{An Alternative to Kaluza}

\author{W.-D. R. Stein}
\address{IFP, Technical University Dresden, D-01062 Dresden, Germany}

\ead{stein@physik.tu-dresden.de}

\date{\today}

\begin{abstract}
In Kaluza derived theories the electromagnetic potential is interpreted as a part of the metric in a higher dimensional theory of gravity. Here we present a more Yang-Mills like unification of classical electromagnetism and gravity within five dimensions, where the electromagnetic potential is related to the connecion and the field strength to the curvature. In addition some aspects of a quantum theory can be described in the context of the Gauss-Bonnet-Chern theorem. This approach offers new coupling mechanisms between gravity and electromagnetism.
\end{abstract}

\pacs{03.50.De, 04.50.Cd, 04.50.Kd}

\submitto{\CQG}

\maketitle

\section{Introduction}

The unification of classical electrodynamics and gravitation
within higher-dimensional space-times has a long tradition since
the discoveries of Nordström \cite{Nordstrom1914a}
and Kaluza \cite{Kaluza1919a} together with the idea of compactification of
Klein \cite{Klein1926a}. 
Kaluzas approach is based on the identification of the additional part of the five-dimensional metric with the electromagnetic potential. The field strength therefore is part of the five-dimensional connections. In contrast, the quantum field theories are more Yang-Mills like, where the field strength is related to a curvature and the potential to a connection. Here we discuss an approach, where classical electromagnetism is included in a five-dimensional theory of gravity and the field strength is related to the curvature.

Classical electrodynamics in four-dimensional space-times can be characterized by the axiomatic approach \cite{Hehl2003a}. The charge current 3-form $J$ fulfills a conservation law and the electromagnetic excitation 2-form $H$ is connected to the charge current via
\begin{equation}
d J = 0, \qquad d H = J.
\end{equation}
The field strength 2-form $F$ fulfills a similar conservation law and the electromagnetic potential 1-form $A$ serves as a potential for the electromagnetic field strength
\begin{equation}
d F = 0, \qquad d A = F.
\end{equation}
Assuming a linear space-time-relation, the field strength and the excitation in vaccum are related by 
\begin{equation}
H = \lambda \star F,
\end{equation}
where $ \star $ denotes the hodge-operator and $\lambda$ the vacuum impedance with $\lambda = \sqrt{\epsilon_0 / \mu_0}$.
The action of electromagnetic fields on charged matter is known to be described by the lorenz-force $ f_\alpha$
\begin{equation}
 f_\alpha = ( e_\alpha \interior F ) \wedge J,
\end{equation}
where $e_\alpha$ is the frame related to the coframe $\vartheta^{\alpha}$ by $e_\alpha \interior \vartheta^\beta = \delta^\beta_\alpha$.
From the lorenz-force the definition of the energy-momentum 3-form $\Sigma_\alpha$ is possible
\begin{equation}
\Sigma_\alpha = \frac{1}{2}[ (e_\alpha \interior F ) \wedge H - (e_\alpha \interior H ) \wedge F].
\end{equation}
These equations characterize the structure of electrodynamics and we will show, that this structure, translated to a five-dimensional space-time, is included in the structure of a nearly arbitrary five-dimensional riemannian manifold.

The gravitational interaction in four-dimensional space-times is well described by the theory of general relativity. Let $\Gamma_\alpha{}^\beta$ be the connection 1-form, we define the curvature 2-form $R_\alpha{}^\beta$ by
\begin{equation}
R_\alpha{}^\beta = d \Gamma_\alpha{}^\beta + \Gamma_\sigma{}^\beta \wedge \Gamma_\alpha{}^\sigma.
\end{equation}
The field equations for a four-dimensional manifold are
\begin{equation}
\frac{1}{2}\eta_{\alpha\beta\gamma} \wedge R^{\beta\gamma} = \frac{8 \pi G}{c^3} \Sigma_\alpha,
\end{equation}
where $\eta_{\alpha\beta\gamma}$ is a 1-form defined by 
\begin{equation}
\eta^{\alpha_1 ... \alpha_p} := \star (\vartheta^{\alpha_1} \wedge ... \wedge \vartheta^{\alpha_p}),
\end{equation}
$G$ is Newtons gravitational constant and $c$ is the speed of light.

Charge is countable. Magnetic flux lines are countable. Both fields fulfill conservation laws. Here we will show, that both fields are expressed by the euler-form of the Gauss-Bonnet-Chern theorem of submanifolds of the underlying space-time.

Charge is countable, in other words it exists only in an integer
multiple of the electric charge. This can formulated as an
integral law
\begin{equation}
\int_\Sigma \rho = n e , \quad n \in \mathbb{Z},
\end{equation}
where the boundary term is assumed to vanish.

The same law can be formulated for the magnetic flux in superconducting materials, but here we have to draw more attention on the boundary. In superconducting materials, the boundary $\Sigma_2$ can be chosen to be within the superconducting phase and we have for the magnetic flux 
\begin{equation}
\int_{\Sigma_2} \mathcal{B} = n \Phi_0, \quad n \in \mathbb{Z},
\end{equation}
where $\mathcal{B}$ is the flux-density and $\Phi_0 =
\frac{h}{2e}$ the flux-quantum. In the following sections we will
see, that we can find a $J$ in terms of geometric quantities which
is closed and when integrating the horizontal part (with respect to a time-like vectorfield) over a space-like submanifold yields
an integer multiple of the elementary charge $e$. The field strength $F$ can also be related to
geometric quantities, is closed and the integral over the horizontal part of $F$ gives in the case of a vanishing boundary term an integer multiple of the flux-quantum $\Phi_0$. The linear space-time-relation will be discussed to be a result of some approximations.
  
The idea consists of the use of the Gauss-Bonnet-Chern (GBC) theorem
for manifolds with boundary \cite{Chern1999a,Eguchi1980a,Gilkey1995a}. The GBC-theorem states, that the integral of the euler-form over a n-dimensional space-like manifold $\Sigma_n$ plus an integral over the boundary $\partial \Sigma_n$ in the case of even $n$ is equal to the euler-characteritic $\chi$\begin{equation}
\int_{\Sigma_n} E_n + \int_{\partial \Sigma_n} Q_n =  \chi
(\Sigma_n, \partial \Sigma_n ),
\end{equation}
where $E_n$ is the euler-form of even dimensional manifolds
\begin{equation}
E_n = \frac{(-1)^{n/2}}{2^n\pi^{n/2} (n/2)!} \eta_{\mu_1 ... \mu_n} R^{\mu_1\mu_2} \wedge ... \wedge R^{\mu_{n-1}\mu_n},
\end{equation}
and $Q_n$ the boundary form.
Applying a time derivative along a time-like vectorfield (neglegting the extra term from the Leibnitz rule), we result in
\begin{equation}\label{exp:conservation}
\frac{d}{d t} \int_{\Sigma_n} E_n + \int_{\partial \Sigma_n}
\frac{\partial}{\partial t} Q_n  = 0,
\end{equation}
which has the form of a conservation law.

\section {Charge current}

Now we start with defining the charge current and the corresponding excitation in terms of geometric quantities, where the charge current fulfills the conservation law and the excitation serves at this point as a potential for the charge current:
\begin{equation}\label{exp:jcon}
d J = 0 \qquad \qquad d H = J.
\end{equation}
In addition the horizontal part of the charge current with respect to a time-like vectorfield should reduce in a special coordinate system to the euler-form  of the Gauss-Bonnet-Chern theorem for four-dimensional manifolds.

The expressions defined as follows for $J$ and $H$ fulfill the equations (\ref{exp:jcon}) :
\begin{equation}\label{exp:j}
J = -\frac{e}{32 \pi^2} \eta_{\alpha\beta\gamma\delta\sigma} (
R^{\alpha\beta} + D u^\alpha \wedge Du^\beta ) \wedge (R^{\gamma\delta} + D u^\gamma
\wedge Du^\delta ) u^\sigma
\end{equation}
and
\begin{equation}\label{exp:h}
 H =  \frac{e}{128 \pi^2} \eta_{\alpha\beta\gamma\delta\sigma}
( R^{\alpha\beta} + D u^\alpha \wedge Du^\beta  - \frac{2}{3} Dw^\alpha \wedge
Dw^\beta) \wedge Dw^\gamma w^\delta u^\sigma,
\end{equation}
where $u^\mu$ is a time-like vectorfield and $w^\mu$ a vectorfield orthonormal to  $u^\mu$. 
As later will be seen, $w^\mu$ is defined by the choice of a space-time-relation. The time-like vectorfield $u^\mu$ will serve as the velocity vectorfield of the charge distribution as in the space-like submanifold orthogonal to $u^\mu$, the horizontal part of $J$ will reduce to the euler-form for a four-dimensional manifold. To make sure, that there always exists a time-like vectorfield $u^\mu$ with the above properties, we need to apply the condition
\begin{equation}\label{exp:condition-J}
u^\mu e_\mu \interior J = 0.
\end{equation}
In the case, that this condition is not satisfied, there are the following cases. First: the solution gives for $u^\mu$ a space-like or light-like vectorfield. This case would be interesting to be discussed in the context of dark matter as maxwells equations are still valid, but the charge distribution loses the property of countability. Second: there exists no solution to the condition above. In this case there exists a vectorfield $q^\mu$ different from $u^\mu$ satisfying the condition
\begin{equation}
q^\mu e_\mu \interior J = 0.
\end{equation}
This case is interesting so far, as it would imply, that in the space-like submanifold orthogonal to $u^\mu$ charge is still countable, but in addition there exists a non-vanishing current in the coordinate system, where charge is countable.

Let now again $u^\mu$ be a time-like vectorfield with the condition (\ref{exp:condition-J}) fulfilled. The space-time decomposition with respect to $u^\mu$ of the charge current and its excitation is
\begin{equation}
J = j \wedge d \sigma -  \rho, \qquad H = \mathcal{H} \wedge d \sigma + D,
\end{equation}
where $d \sigma$ is the 1-form dual to the vectorfield $u^\mu$ with $u \interior d \sigma = 1$. In the four-dimensional submanifold $\Sigma_\sigma$ orthogonal to the vectorfield $u^\mu$ the charge density $\rho$ becomes
\begin{equation}
\rho = \frac{e}{32 \pi^2}  \eta^{\Sigma_\sigma}_{abcd} \;
 R_{\Sigma_\sigma}^{ab} \wedge R_{\Sigma_\sigma}^{cd},
\end{equation}
where the latin indices run over the index range of the four-dimensional submanifold $\Sigma_\sigma$. This expression equals the product of the elementary charge $e$ and the euler-form of four-dimensional manifolds. When shifting the boundary of the integration volume to infinity, the boundary term of the GBC-theorem vanishes and the integration results in an integer multiple of the elementary charge $e$.

\section{Field strength}

Similar to the description of the charge current and its excitation, we can find some geometric  expressions for the field strength and its potential, which fulfill the following equations
\begin{equation}
d F = 0 \qquad \qquad d A = F,
\end{equation}
and the horizontal part of $F$ reduces in a two-dimensional space-like hyper-surface to the euler-form of the GBC-theorem. 

Choosing three orthonormal vectorfields, one time-like vectorfield $v^\mu$ and two space-like vectorfields $n^\mu$ and $z^\mu$ we find
\begin{equation}\label{exp:f}
F =  -\frac{h}{8\pi e} \eta_{\alpha\beta\gamma\delta\sigma} ( R^{\alpha\beta} + D v^\alpha \wedge Dv^\beta  + D z^\alpha \wedge Dz^\beta + D n^\alpha \wedge Dn^\beta ) z^\gamma n^\delta v^\sigma
\end{equation}
and
\begin{equation}
A =  \frac{h}{16\pi e} \eta_{\alpha\beta\gamma\delta\sigma}
Dx^\alpha x^\beta z^\gamma n^\delta v^\sigma,
\end{equation}
where $x^\mu$ is an arbitrary space-like vectorfield within the chosen two-dimensional hyper-surface orthogonal to $v^\mu,n^\mu$ and $z^\mu$. The prefactors have been chosen to result for the integration of the horizontal part of the field strength in integer multiples of the flux quantum $\Phi_0 = \frac{h}{2 e}$.  As  $x^\mu$ is arbitrary also the choice  $\tilde{x}^\mu =  x^\mu \cos{\varphi} +  y^\mu \sin{\varphi} $, where $y^\mu$ is orhonormal to $v^\mu,n^\mu, z^\mu$ and $ x^\mu$, is valid. The remaining freedom of choice of the phase $\varphi$ is represented by the gauge-group $SO(2)$, which is isomorphic to the established gauge-group $U(1)$ of electrodynamics.

The space-time decomposition of the field strength and the potential results in
\begin{equation}
F = E \wedge d \sigma +  B \qquad A =\phi \wedge d \sigma + \mathcal{A}.
\end{equation}
As the hyper-surface where the magnetic flux is countable is orthogonal to the vectorfields $n^\mu$ and $z^\mu$, the following conditions need to be fulfilled to assure the countability of magnetic flux
\begin{equation}
n^\mu e_\mu \interior B = 0 \qquad z^\mu e_\mu \interior B = 0.
\end{equation}
In the space-like two-dimensional submanifold $\Sigma_2$ orthogonal to the vectorfields $v^\mu,n^\mu$ and $z^\mu$ the horizontal part of the field strength $F$ reduces to the euler-term of the GBC-theorem times the flux-quantum
\begin{equation}
B_{\Sigma_2} = -\frac{h}{8\pi e}  \eta^{\Sigma_2}_{ab} \;
 R_{\Sigma_2}^{ab},
\end{equation}
where the latin indices are restricted to the index range of the hyper-surface $\Sigma_2$.

Shifting the boundary to infinity, the boundary term vanishes and the integral over $\Sigma_2$ results in an integer multiple of the flux-quantum $\Phi_0 = \frac{h}{2 e}$. The time-derivative of the boundary-term is related to the electric field $E$ on the boundary (see equation (\ref{exp:conservation})), which will vanish in superconducting materials. However in superconducting materials not only the time-derivative, but also the boundary-term itself seems to vanish, resulting for the integration of $B$ in an integer multiple of the flux-quantum.

\section{Space-Time-Relation}

The relationship between $F$ and $H$ is called the
space-time-relation. Several different relations have been
discussed in literature (see e.g. reference \cite{Hehl2003a}). Here we show, that under
some assumptions a linear relationship
\begin{equation}
H = \lambda \star F  
\end{equation}
can be found.
First assuming, that the square terms of the covariant differentials $Du^\mu\wedge D u^\nu$, $Dn^\mu\wedge D n^\nu$ and $Dz^\mu\wedge D z^\nu$ of the
vector fields $u^\mu$, $n^\mu$ and $z^\mu$ are small compared to the curvature
term, we result for the field strength (\ref{exp:f}) in
\begin{equation}\label{exp:f-1}
F \approx -\frac{h}{8\pi e}\eta_{\alpha\beta\gamma\delta\sigma} R^{\alpha\beta} z^\gamma
n^\delta v^\sigma
\end{equation}
and for the excitation field (\ref{exp:h}) in
\begin{equation}
H \approx  \frac{e}{128\pi^2}\eta_{\alpha\beta\gamma\delta\sigma} R^{\alpha\beta} \wedge D
w^\gamma w^\delta u^\sigma.
\end{equation}
Further we need to evaluate some formulas for the hodge-duals of the
curvature. Denoting the non-decomposable parts of the anti-symmetric part of the curvature by ${}^{(i)} R_{\mu\nu}$ the hogde-duals can be evaluated in analogy to the four-dimensional case (\cite{Heinicke2005a, McCrea1992a}). For the equality
\begin{equation}
e^\nu \interior e^\mu \interior (\eta^{\alpha\beta} \wedge
{}^{(i)} R_{\mu\nu}) = e^\nu \interior e^\mu \interior (\star
{}^{(i)} R_{\mu\nu} \wedge \vartheta^\alpha \wedge
\vartheta^\beta)
\end{equation}
we find, that for the hodge-dual of the irreducible parts of the curvature the following can be proofed after
some algebra
\begin{equation}
 {\star} {}^{(i)} R^{\alpha\beta} = \frac{a_i}{2} \eta^{\alpha\beta\mu\nu} \wedge {}^{(i)}
 R_{\mu\nu}
\end{equation}
with
\begin{equation}
a_i = \left( 1, -1,  1,  -\frac{n-2}{2(n-3)},  n-3,   \frac{2}{(n-2)!}  \right).
\end{equation}
In a five dimensional riemannian manifold only the values $a_1 =
1, a_4 = -3/4, a_6 = 1/3$ are of interest.

The hodge-dual of $F$ (equation (\ref{exp:f-1})) in this case takes the form
\begin{equation}
\star F \approx -\frac{h}{8\pi e} \frac{1}{2} \eta_{\alpha\beta\gamma\delta\sigma}
\eta^{\alpha\beta\mu\nu}  \wedge ({}^{(1)} R_{\mu\nu} -
\frac{3}{4}{}^{(4)} R_{\mu\nu}  + \frac{1}{3}{}^{(6)} R_{\mu\nu}
)  z^\gamma n^\delta v^\sigma.
\end{equation}
The summation over $\alpha$ and $\beta$ results in a simpler equation
\begin{equation}
\star F \approx  -\frac{h}{8\pi e} \delta_{\gamma\delta\sigma}^{\mu\nu\kappa}
\vartheta_\kappa \wedge ({}^{(1)} R_{\mu\nu} - \frac{3}{4}{}^{(4)}
R_{\mu\nu} + \frac{1}{3}{}^{(6)} R_{\mu\nu} ) z^\gamma n^\delta
v^\sigma.
\end{equation}
A further reasonable assumption comes from the field equations of
gravity, where the Ricci-part ${}^{(4)}R_{\mu\nu}$ and the trace-part ${}^{(6)}R_{\mu\nu}$ are related to energy-momentum. For low energy-momentum density the Weyl-part ${}^{(1)}R_{\mu\nu}$ becomes the leading term in this expression and  we result in
\begin{equation}
\star F \approx -\frac{h}{8\pi e} \delta_{\gamma\delta\sigma}^{\mu\nu\kappa}
\vartheta_\kappa \wedge {}^{(1)} R_{\mu\nu}z^\gamma n^\delta
v^\sigma.
\end{equation}
Comparing with the approximation of the excitation field
\begin{equation}
H \approx  \frac{e}{128\pi^2}\eta_{\alpha\beta\gamma\delta\sigma} {}^{(1)} R^{\alpha\beta} \wedge D
w^\gamma w^\delta u^\sigma,
\end{equation}
one needs to solve the the equation
\begin{equation}
-\lambda \frac{h}{8\pi e} \delta_{\gamma\delta\sigma}^{\mu\nu\kappa}
\vartheta_\kappa  z^\gamma n^\delta
v^\sigma =   \frac{e}{128\pi^2} \eta^{\mu\nu\gamma\delta\sigma}  D w_\gamma w_\delta u_\sigma.
\end{equation}
We turn to a coordinate system, where the only non-vanishing parts of the vectorfields $z^\mu, n^\mu$ and $v^\mu$ are $z^3, n^4$ and $v^0$. We find with $\lambda = \sqrt{\epsilon_0 / \mu_0} = e^2 / (2 h \alpha_f)$ (where $\alpha_f$ is Sommerfeld's fine structure constant)
\begin{equation}
-\frac{8 \pi }{ \alpha_f} \delta_{340}^{\mu\nu\kappa}
\vartheta_\kappa   =   \eta^{\mu\nu\gamma\delta\sigma}  D w_\gamma w_\delta u_\sigma.
\end{equation}
A possible solution would result by assuming the only non-vanishing parts of $w^\mu$ and $Dw^\mu$ to be given by $\mu = 1,2$. In this case we have
\begin{equation}
-\frac{8 \pi }{ \alpha_f} \delta_{340}^{\mu\nu\kappa}\vartheta_\kappa =  \delta_{340}^{\mu\nu\kappa}u_\kappa (\delta^{\alpha\beta}_{12} D w_\alpha w_\beta).
\end{equation}
The solution results for $u^\mu= 0, \mu=1,2$ in
\begin{equation}
u^\mu (e_\mu \interior \Gamma_1{}^2)  = -\frac{8 \pi}{\alpha_f} .
\end{equation}
This solution describes a vectorfield $w^\mu$, which is rotating in the form of a spiral along $u^\mu$ within the two-dimensional hyper-planes orthogonal to $z^\mu, n^\mu$ and $v^\mu$ with a velocity proportional to the inverse of the fine structure constant.

\section{Energy-Momentum}

The interaction of the electromagnetic fields with charged matter is described by the lorenz-force
\begin{equation}
f_\alpha = (e_\alpha \interior F) \wedge J,
\end{equation}
where $F$ and $J$ have already been defined in terms of geometrical expressions.
From the lorenz-force we can define the energy-momentum by 
\begin{equation}
f_\alpha  =: D \Sigma_\alpha + X_\alpha,
\end{equation}
where the energy-momentum $\Sigma_\alpha$ in five-dimensional manifolds is (note the change in sign compared to the four-dimensional case)
\begin{equation}
\Sigma_\alpha = \frac{1}{2}[(e_\alpha \interior F) \wedge H + (e_\alpha \interior H) \wedge F]
\end{equation}
and the extra force $X_\alpha$ is given by
\begin{equation}
X_\alpha = -\frac{1}{2} ( F \wedge \mathcal{L}_\alpha H - H \wedge \mathcal{L}_\alpha F),
\end{equation}
where $\mathcal{L}_\alpha$ is the covariant Lie-derivative. Here again the energy-momentum and the extra force are defined by geometrical expressions trough $F$ and $H$. The covariant Lie-derivative of a p-form $\Psi$ is
\begin{equation}
\mathcal{L}_\alpha \Psi = \frac{1}{p!}(D_\alpha \Psi_{\alpha_1 ... \alpha_p}) \vartheta^{\alpha_1}\wedge ... \wedge \vartheta^{\alpha_p}.
\end{equation}
The extra force can be evaluated similar to the four-dimensional case (\cite{Hehl2003a}). We define
\begin{equation}
\hat{\mathcal{H}} ^{\alpha\beta}:=\frac{1}{6} \eta^{\alpha\beta\mu\nu\kappa} H_{\mu\nu\kappa},
\end{equation}
where $H_{\mu\nu\kappa}$ are the components of the 3-form $H$. Further we assume a linear space-time-relation
\begin{equation}
\hat{\mathcal{H}} ^{\alpha\beta}=\frac{1}{2} \chi^{\alpha\beta\mu\nu} F_{\mu\nu},\end{equation}
where $F_{\mu\nu}$ are the components of the field strength 2-form $F$ and $\chi^{\alpha\beta\mu\nu}$ is the linear permittivity. Combining these two equations we find for the extra force
\begin{equation}
X_\beta =  -\frac{\eta}{8} (D_\beta \chi^{\mu\nu\alpha\kappa} ) F_{\mu\nu} F_{\alpha\kappa}.
\end{equation}
In the case of vanishing covariant derivative $D_\beta \chi^{\mu\nu\alpha\kappa}$, the equation for the lorenz-force reduces to
\begin{equation}
 f_\alpha = D \Sigma_\alpha,
\end{equation}
which is for field configurations with vanishing lorenz-force a conservation law for the energy-momentum density.

\section{Summary}

By using the Gauss-Bonnet-Chern theorem for even-dimensional
submanifolds of a five-dimensional manifold, we could show, that
the structure of classical electrodynamics can be described in
terms of geometrical expressions in a total different way, as for
example in Kaluza-Klein theories. The conservation laws for the
charge and for the magnetic flux hold exactly. Under some
assumptions a linear space-time-relation coupling the excitation
and the field strength can be found. The energy-momentum will be conserved  in the absence of the extra force and for field configurations, where the lorenz-force vanishes. As the electromagnetic fields depend on the curvature, new coupling mechanisms between gravity and electromagnetism are possible.


\section*{References}


\begin{thebibliography}{10}




\bibitem{Nordstrom1914a}
G.~Nordström {\sl et al.}, Phys.~Z.  {\bf 15}, 504 (1914).

\bibitem{Kaluza1919a}
Th.~Kaluza {\sl et al.}, Sitzungsber.~Preuss.~Akad.~Wiss. Phys.~Math. Klasse 996 (1921).

\bibitem{Klein1926a}
O.~Klein {\sl et al.}, Zs.~f.~Physik {\bf 37}, 895 (1926).




\bibitem{Hehl2003a}
F.-W.~Hehl {\sl et al.}, {\sl Foundations of classical electrodynamics}. Birkhäuser, 2003.

\bibitem{Chern1999a}
S.~S.~Chern {\sl et al.}, {\sl Lectures on Differential Geometry}. World Scientific, 1999.

\bibitem{Eguchi1980a}
T.~Eguchi {\sl et al.}, Phys.~Reports {\bf 66}, 213 (1980).


\bibitem{Gilkey1995a}
P.~B.~Gilkey, {\sl Invariance theory, the Heat Equation and the Atiyah-Singer Index Theorem}. CRC Press, 1995.

\bibitem{Heinicke2005a}
C.~Heinicke, {\sl Exact solutions in Einstein's theory and beyond}. PhD-Thesis, 2005.

\bibitem{McCrea1992a}
J.~D.~McCrea {\sl et al.}, Class.~Quantum~Grav. {\bf 9}, 553 (1992).


\end{thebibliography}
\end{document}